\documentclass{article}

\usepackage{amsmath}
\usepackage{amssymb}
\usepackage{braket}
\usepackage{booktabs}
\usepackage{tabularx}
\usepackage{geometry}
\usepackage{graphicx}
\usepackage{multirow}
\usepackage{subfigure}
\usepackage{threeparttable}
\usepackage{xcolor}
\usepackage{array}
\newcolumntype{C}[1]{>{\centering\arraybackslash}m{#1}}

\usepackage[normalem]{ulem}

\title{Recent Advances in Quantum Architecture Search}
\author{Haozhen Situ$^{1}$, Zhimin He$^{2}$, Lvzhou Li$^{3,}$\thanks{Corresponding author. lilvzh@mail.sysu.edu.cn}\\
\small $^{1}$College of Mathematics and Informatics, South China Agricultural University, Guangzhou 510642, China\\
\small $^{2}$School of Electronic and Information Engineering, Foshan University, Foshan, 528000, China\\
\small $^{3}$Institute of Quantum Computing and Software, School of Computer Science and Engineering, \\
\small Sun Yat-sen University, Guangzhou, 510006, China
}

\date{}

\begin{document}
\maketitle
\begin{abstract}
Variational quantum algorithms (VQAs) constitute a prominent paradigm for exploring the capabilities of near-term quantum computers. As the effectiveness of VQAs depends heavily on the design of variational quantum circuits, Quantum Architecture Search (QAS) has emerged as a critical research area to automate the discovery of high-performing circuit structures. This paper reviews key advancements in current research on QAS, including its core concepts, representative methodologies, applications, and emerging trends. The content is structured to ensure broad accessibility for a diverse audience of researchers while preserving the core principles of complex methodologies. In addition, we discuss remaining challenges and suggest potential research directions to offer perspectives on future exploration in this rapidly evolving field.
\end{abstract}

\section{Introduction}
\label{Sect:intro}

One of the main motivations for work in quantum information science is the prospect of
fast quantum algorithms to solve important computational problems \cite{montanaro2016quantum,zhang2022brief}. Since  the first quantum algorithm, Deutsch’s algorithm \cite{Deutsch1985}, was proposed in 1985,  significant efforts have been made in quantum algorithm design over the past forty years. Beyond the well-known quantum speedups in foundational problems such as factorization \cite{Shor1994}, unordered database search \cite{Grover1996}, and solving linear equations \cite{HHL2009}, quantum algorithms have also demonstrated advantages in solving  combinatorial optimization problems \cite{jordan2025optimization}, NP search problems \cite{YamakawaZ22}, the welded tree  problem \cite{ChildsCDFGS03,jeffery2023multidimensional,li2024recovering-a,li2024recovering-b},  spatial search problems \cite{ quadratic_20, quadratic_22, li2025quantum,li2026deterministic,wang2025unifying}, geometric problems \cite{li2025unbounded, nghiem2025quantum, andrejevs2026quantum},  string learning problems \cite{xu2025provable, qi2026quantum, xu2023quantum, AmbainisM14, CleveIGNTTY12}, problems with abstract structures (e.g., groups  \cite{childs2010quantum, kuperberg2005subexponential,  watrous2001quantum,  magniez2007quantum, ye2022deterministic}, matroids \cite{huang2024quantum-a,huang2024quantum-b,huang2026quantum}), and others. 

Designing quantum algorithms  is inherently challenging because it requires specialized knowledge of the problem structure and the ability to exploit nonintuitive quantum mechanical effects to achieve an advantage over  classical algorithms.  Instead of specifying the exact procedure of a quantum algorithm, variational quantum algorithms (VQAs) adopt variational quantum circuits (also known as parameterized quantum circuits or ansatzes) as learning models and use classical optimization methods to train the circuit parameters by minimizing a task-dependent loss function. In the current noisy intermediate-scale quantum era, VQAs have become a prominent paradigm for exploring the capabilities of near-term quantum computers. Typical examples include the variational quantum eigensolver (VQE) \cite{peruzzo2014variational} for ground state energy estimation, the quantum approximate optimization algorithm (QAOA) \cite{farhi2014quantum} for combinatorial optimization, and quantum classifiers \cite{mitarai2018quantum}. For a comprehensive overview of VQAs, we refer the reader to \cite{cerezo2021variational,wang2026review}. 

The effectiveness of VQAs depends heavily on the design of the variational quantum circuit. Traditional approaches rely on fixed circuit structures, which often contain redundant parameters and excessive depth that complicate optimization. These issues also increase noise accumulation and degrade performance. In addition, hardware-agnostic designs usually require a compilation step to adapt circuits to the qubit connectivity and native gate sets of specific quantum devices. This step can significantly increase gate count and circuit depth, which further amplifies the impact of noise. As a result, end-to-end methodologies have become increasingly important because they enable the automatic discovery of circuit structures that are tailored to both the target task and the hardware. 

A machine-learning approach was proposed for the automated discovery of short-depth quantum algorithms \cite{cincio2018learning}. This method utilizes simulated annealing to identify novel quantum circuits for computing state overlap $\mathrm{Tr}(\rho\sigma)$ that surpass the standard Swap Test in efficiency. An approach called VAns (Variable Ansatz) was introduced to dynamically construct quantum circuits during the optimization process \cite{bilkis2023semi}. It also utilizes simulated annealing and introduces rules for the targeted addition and removal of quantum gates. The adaptive derivative-assembled pseudo-Trotter ansatz variational quantum eigensolver (ADAPT-VQE) algorithm \cite{grimsley2019adaptive} builds quantum circuits by iteratively selecting operators from a predefined pool based on the largest energy gradient contribution. The adaptive diversity-based quantum architecture search (ADQAS) algorithm \cite{huang2024adaptive} employs an iterative block-addition strategy together with diversity-guided sampling to prune the search space and ensure structural variety among candidate circuits. 

In recent years, research on the automated design of quantum circuit structures has gradually developed into the research area now commonly referred to as Quantum Architecture Search (QAS). This development has been inspired in part by neural architecture search (NAS) \cite{salmani2025systematic}, which automates the design of neural network architectures, and the term QAS reflects this close conceptual connection to NAS. QAS shares with NAS the general objective of automating architecture design and builds on several methodological ideas that are also central to NAS. At the same time, QAS must address the characteristics and constraints specific to quantum circuits. QAS can also be viewed in the broader context of the growing use of artificial intelligence to support and accelerate scientific discovery across different disciplines~\cite{shan2026artificial,li2026AI}.

While a substantial body of QAS research focuses on parameterized circuits for VQAs, the scope of QAS considered in this review also includes the search for circuits composed entirely of gates without tunable parameters. For example, Ref.~\cite{zhang2020topological} uses deep reinforcement learning to compile an arbitrary single-qubit unitary into a sequence of elementary gates, while Ref.~\cite{kuo2021quantum} employs deep reinforcement learning to design quantum circuits for preparing Bell and GHZ states. In both studies, the elementary gates used to construct the circuits contain no tunable parameters.

QAS offers significant advantages by automating circuit design for a wide range of target tasks and by directly incorporating hardware constraints in an end-to-end manner. By searching for circuits that respect hardware topology and native gate sets from the outset, QAS can avoid the overhead associated with standard compilation and produce more compact and efficient circuits. A concise comparison between traditional manual design and QAS is provided in Table~\ref{tab:qas_vs_manual}. Overall, QAS shifts circuit design from an expert-driven process to an automated search paradigm, enabling the discovery of more compact circuit structures while better accounting for hardware constraints and noise characteristics.

\begin{table}[h]
\centering
\caption{Comparison between manual design and QAS}
\label{tab:qas_vs_manual}
\begin{tabular}{|c|c|c|}
\hline
 & \textbf{Manual Design} & \textbf{QAS} \\ \hline
\textbf{Methodology} & expert-driven & automated, task-agnostic search \\ \hline
\textbf{Circuit Structure} & fixed, often redundant & adaptive and compact \\ \hline
\textbf{Hardware Adaptation} & compilation overhead & hardware-aware design \\ \hline
\textbf{Noise Awareness} & performance degradation & noise-robust optimization \\ \hline
\end{tabular}
\end{table}

\begin{figure}[htbp]
    \centering
    \includegraphics[width=0.95\textwidth]{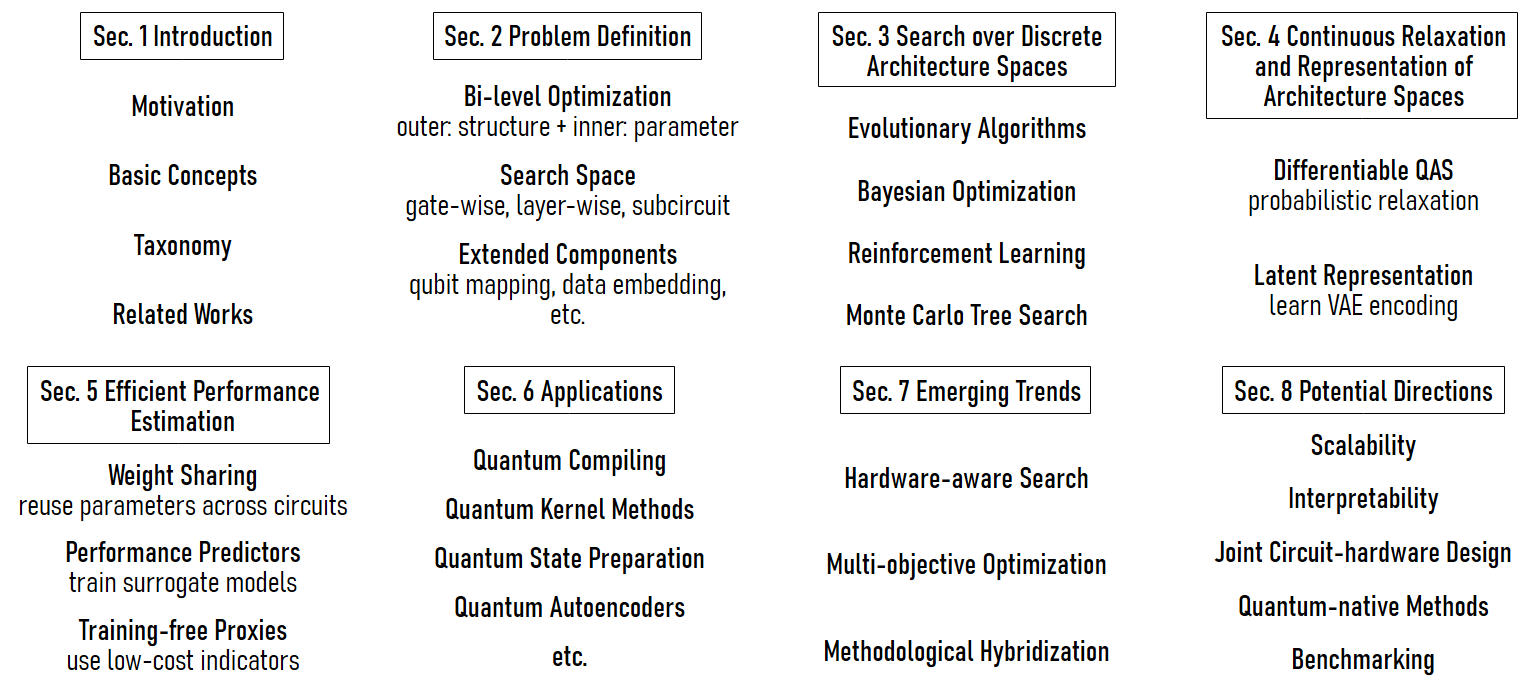} 
    \caption{Schematic taxonomy and organization of this review.}
    \label{fig:organization} 
\end{figure}

This paper aims to provide a structured perspective on the rapidly evolving field of QAS. The presentation is intended to remain accessible to a broad research audience while preserving the essential technical ideas. For clarity, Figure \ref{fig:organization} presents the taxonomy proposed in this paper, which provides a unified perspective for organizing and understanding the rapidly growing body of QAS research. Specifically, we categorize existing QAS research into several key pillars, including search over discrete architecture spaces, continuous relaxation and representation of architecture spaces, efficient performance estimation, and application domains. Guided by this taxonomy, Section \ref{Sect:problem} provides a formal definition of the QAS problem and reviews common search space designs together with extended search components beyond circuit structures. Section \ref{Sect:discrete} surveys representative search strategies over discrete spaces of quantum circuit structures, including evolutionary algorithms, Bayesian optimization, reinforcement learning, and Monte Carlo tree search. 
Section \ref{Sect:d2c} examines two approaches that enable the search over continuous structure spaces through relaxation or learned representation.
Section \ref{Sect:efficient} reviews methods for efficient performance estimation without fully training candidate quantum circuits, including weight sharing, performance predictors, and training-free proxies. 
Section \ref{Sect:application} summarizes representative application domains of QAS. 
Section \ref{Sect:trends} discusses several emerging trends in the field of QAS.
Finally, Section \ref{Sect:conclusion} outlines remaining challenges and discusses potential research directions for future work. 

It should be noted that the categories are not intended to be mutually exclusive. Rather, they represent several complementary dimensions of QAS. A single work may therefore be associated with more than one category. For example, a QAS work may use evolutionary algorithms to explore the search space and adopt efficient performance estimation methods to reduce the cost of evaluating candidate circuits. We make every effort to place each work under the topic where it most appropriately illustrates the corresponding aspect of QAS, with cross-references provided in other relevant contexts.

In this review, we focus on the major paradigms that have been widely explored in the QAS literature. The above-mentioned automated circuit construction approaches based on simulated annealing \cite{cincio2018learning,bilkis2023semi} and adaptive methods \cite{grimsley2019adaptive,huang2024adaptive} can also be regarded as forms of QAS, since they share the common goal of automating the design of quantum circuit structures. Nevertheless, these circuit construction approaches have not been explored as extensively as the major search paradigms examined in dedicated sections of this review. We therefore do not treat them as separate categories.

Several reviews have discussed aspects of QAS and related quantum circuit design techniques. 
A brief survey of QAS focuses on research works based on evolutionary algorithms, reinforcement learning, and differentiable QAS \cite{zhu2023brief}, but it does not provide a systematic organization of the field.
Another review centers on various VQA improvement approaches \cite{qin2023review}, including several QAS algorithms, but does not systematically examine QAS as an independent automated circuit design problem.
An additional survey \cite{martyniuk2024quantum} covers a broader range of QAS paradigms and proposes a taxonomy based on different components of QAS methods. 
Compared with these surveys, our review goes beyond summarizing individual QAS methods by providing a systematic, in-depth analysis of QAS, including: (1) the design rationales behind various methods, (2) the advantages, limitations, and trade-offs associated with different paradigms, (3) the comparisons and relationships among QAS approaches, (4) representative applications, and (5) emerging trends in the field. Through this analysis, our review provides valuable insights into the evolution and future development of QAS.

To facilitate understanding for readers from different research backgrounds, Table \ref{tab:terminology} summarizes several key terms used throughout the text.

\begin{table*}[t]
\centering
\begingroup

\caption{Glossary of key terms}
\label{tab:terminology}

\renewcommand{\arraystretch}{1.15}
\setlength{\tabcolsep}{5pt}
\footnotesize

\begin{tabularx}{\textwidth}{
    >{\raggedright\arraybackslash}p{0.32\textwidth}
    >{\raggedright\arraybackslash}X
}
\toprule
\textbf{Term} & \textbf{Definition} \\
\midrule

Variational Quantum Algorithm (VQA)
& A hybrid quantum--classical algorithm in which a parameterized quantum circuit is evaluated on a quantum device or simulator, while its parameters are optimized using a classical algorithm. \\

Variational Quantum Circuit / Parameterized Quantum Circuit / Circuit Ansatz
& A quantum circuit with tunable gate parameters that serves as the trainable model in a variational quantum algorithm, analogous to a neural network in classical machine learning. It can also be viewed as a circuit-based realization of a variational ansatz in physics. \\

Circuit Training
& The optimization of the tunable parameters of a fixed quantum circuit structure with respect to a task-dependent objective function. In QAS, it generally corresponds to the inner parameter optimization process. \\

Quantum Architecture Search (QAS)
& The automated process of identifying quantum circuit structures that achieve good performance on a target task, typically within a predefined search space and under specified hardware or resource constraints. \\

Variational Quantum Eigensolver (VQE)
& A variational quantum algorithm that estimates the ground state energy of a Hamiltonian by minimizing its expectation value using a variational quantum circuit. \\

Quantum Approximate Optimization Algorithm (QAOA)
& A variational quantum algorithm for combinatorial optimization that reformulates an objective minimization problem as the task of finding the ground state of a corresponding problem Hamiltonian.
\\

Quantum Classifier
& A quantum machine learning model that uses a quantum circuit to process input data and predicts class labels from measurement outcomes. \\

Data Embedding / Feature Encoding
& The process of mapping classical input data into quantum states so that the data can be processed by a quantum circuit. \\

Expressibility
& The ability of a parameterized quantum circuit to represent a broad range of quantum states as its parameters vary. \\

Trainability
& The ability of a variational quantum circuit to be effectively optimized for a given objective, typically influenced by factors such as gradient magnitude, optimization landscape, circuit depth, and noise.
\\

Learnability
& The ability of a variational quantum circuit to achieve the desired performance on a target task through parameter optimization. It depends on both the trainability and expressive capacity of the circuit.
\\

Support Vector Machine (SVM)
& A supervised machine learning method used primarily for classification by identifying a decision boundary that maximizes the margin between different classes.
\\

Autoencoder
& An unsupervised learning model composed of an encoder and a decoder, where the encoder maps input data to a lower-dimensional latent representation and the decoder reconstructs the original input. It is commonly used for dimensionality reduction, feature learning, and data generation.\\

\bottomrule
\end{tabularx}

\endgroup
\end{table*}

\section{Problem Definition}
\label{Sect:problem}

QAS is formally defined as an optimization task directed at identifying the most effective quantum circuit structure $\mathcal{A}$ within a predefined search space $\mathcal{S}$ to minimize an objective function $\mathcal{L}$. Formally, $\mathcal{A}$ denotes the sequence of quantum gates that constitutes the entire circuit, specifying both the type and the placement of each gate as indicated by the respective qubit indices it acts upon. The search space $\mathcal{S}$ is defined as the set of all candidate circuit structures $\mathcal{A}$ that are typically composed of single-qubit and two-qubit gates selected from a predefined gate set $\mathcal{G}$. The gates in $\mathcal{G}$ often include tunable parameters $\theta$ that enable the circuit to be trained. 
Physical qubit coupling constraints of the hardware can be incorporated by ensuring that two-qubit gates are only applied to the pairs of qubits allowed by the qubit connectivity graph. 
Other hardware properties, including gate fidelities, coherence times, and readout errors, can be taken into account when evaluating candidate circuits.
Qubit mapping, the process of assigning the qubits in a quantum circuit to physical qubits on a target device, is also involved in certain QAS research \cite{wang2022quantumnas,situ2024distributed,anagolum2024elivagar,he2025adaptive}.

The objective function $\mathcal{L}$ evaluates the performance on target tasks such as minimizing the expectation value of the Hamiltonian $\langle \phi_{out} | H | \phi_{out} \rangle$. In this formulation, the output state is defined as $| \phi_{out} \rangle = U_{\mathcal{A},\theta} | \phi_{in} \rangle$, where $U_{\mathcal{A},\theta}$ represents the unitary transformation corresponding to the quantum circuit structure $\mathcal{A}$ parameterized by $\boldsymbol{\theta}$. The QAS process is formulated as a bi-level optimization problem where the optimal structure $\mathcal{A}^*$ is found by solving
\begin{align}\label{eq:outer-opt}
\mathcal{A}^* = \arg\min_{\mathcal{A} \in \mathcal{S}} \mathcal{L}(\mathcal{A}, \theta_{\mathcal{A}}^*).
\end{align}
Here, $\theta_{\mathcal{A}}^*$ denotes the optimal set of variational parameters for a candidate structure $\mathcal{A}$ obtained through the inner optimization
\begin{align}\label{eq:inner-opt}
\theta_{\mathcal{A}}^* = \arg\min_{\theta} \mathcal{L}(\mathcal{A}, \theta).
\end{align}
The outer optimization over the structure $\mathcal{A}$ is a discrete optimization problem, while the inner optimization over the parameters $\theta$ is a continuous optimization task, which is widely recognized as the training process for variational quantum circuits. For circuits composed entirely of gates without tunable parameters, Eq. (\ref{eq:outer-opt}) reduces to $\mathcal{A}^* = \arg\min_{\mathcal{A} \in \mathcal{S}} \mathcal{L}(\mathcal{A})$, and the inner optimization in Eq. (\ref{eq:inner-opt}) is not required.

Consider an $N$-qubit system with a gate set defined as $\mathcal{G}=\{R_y, R_z, CNOT\}$. The single-qubit gates $R_y$ and $R_z$ can be independently applied to any of the $N$ qubits, providing $2N$ candidate operations at each position of the gate sequence. The placement of two-qubit $CNOT$ gates is strictly governed by the qubit connectivity graph $G=(V, E)$, where the vertex set $V$ represents the $N$ qubits and the edge set $E$ denotes the available physical couplings. By accounting for the distinct orientations of control and target qubits across these couplings, the number of permissible $CNOT$ placements is $2|E|$. Consequently, the total number of length-$L$ gate sequences is given by $(2N+2|E|)^L$, exhibiting exponential growth relative to the number of gates.

To mitigate the prohibitive scale of the search space, a layer-wise construction strategy may be adopted as an alternative to gate-wise assembly by utilizing predefined layers as the fundamental building blocks of the circuit. For instance, a layer can be defined as a collective operation where $R_y$ or $R_z$ gates are applied to all $N$ qubits simultaneously, or as an entangling block consisting of $CNOT$ gates arranged in a specific pattern with a fixed depth. Assuming the search is restricted to $M$ types of predefined candidate layers, the total number of circuits composed of $L$ layers is $M^L$, which scales exponentially with the number of layers.

The modular definition of layers facilitates some level of transferability from small-scale to large-scale systems, and the resulting layer-wise search space generally exhibits a lower cardinality than the gate-wise approach. In contrast, while gate-wise assembly offers greater structural flexibility and theoretically encompasses the global optimum, it is also more susceptible to generating structurally unbalanced circuits that may lead to poor performance, making the search process highly challenging. Another modular approach, referred to as ``micro search'' \cite{wu2023quantumdarts}, has emerged to strike a balance between fine-grained flexibility and structural regularity. Micro search aims to identify quantum subcircuits that serve as modular building blocks for the larger circuit. Once these subcircuits are discovered, the overall circuit is constructed by assembling them according to predefined compositional rules.

Some works \cite{zhao2024hierarchical,su2025topology} decouple circuit structure optimization into two stages by first determining where gates should be placed and then refining these placement patterns by assigning concrete gate types. This hierarchical \cite{zhao2024hierarchical} or two-phase \cite{su2025topology} design substantially reduces the combinatorial complexity of the search problem. Motivated by the observation that the placement pattern of gates has a more significant impact on performance than the specific choice of gate types, this strategy improves search efficiency while still retaining the ability to discover high-performing circuit structures.

In addition to searching for circuit structures, recent QAS approaches have been extended to optimize other design choices, including feature dimensionality reduction \cite{altares2024autoqml}, data embedding \cite{anagolum2024elivagar}, feature encoding \cite{moretti2025enhanced}, and training settings \cite{situ2025automl}. This joint optimization is naturally compatible with QAS, since these components can be parameterized and incorporated into the search space with relatively low overhead. 

\section{Search over Discrete Architecture Spaces}
\label{Sect:discrete}

Quantum circuit structures are defined by discrete choices, such as gate types and the qubits on which each gate acts. The approaches reviewed in this section operate directly on such discrete spaces using four well-established search strategies: evolutionary algorithms, Bayesian optimization, reinforcement learning, and Monte Carlo tree search. It should be noted that some of these strategies may also be used to optimize continuous gate parameters \cite{yao2020policy,khairy2020learning} instead of conventional gradient-based or derivative-free optimizers, such as Adam and COBYLA, respectively. Nevertheless, the circuit structure itself remains discretely represented and explored. This differs from the approaches introduced in Section \ref{Sect:d2c}, which transform the discrete circuit structure space into a continuous one and perform the search within this continuous space. Therefore, the categorization between this section and Section \ref{Sect:d2c} is based on the representation of the circuit structure space as discrete or continuous, rather than on the specific optimization strategies employed.

\subsection{Evolutionary Algorithms}
\label{Sect:EA} 

Evolutionary algorithms are a class of optimization techniques inspired by mechanisms of biological evolution. The genetic algorithm is a canonical instance, characterized by its use of crossover and mutation operators to evolve a population of candidate solutions. In this process, candidate solutions are represented as individuals within a population, where each individual is encoded as a genome composed of genes that capture the characteristics of the solution. The quality of each individual is evaluated by a fitness function, which quantifies its performance on a given task. New generations are produced through a selection process that favors high-fitness individuals, followed by the application of crossover (recombining genetic material from parent solutions) and mutation (introducing random variations) to promote exploration. Through this iterative process, the population gradually converges toward high-quality solutions, making this approach particularly effective for complex and discrete optimization problems.
  
The Markovian quantum neuroevolution (MQNE) algorithm \cite{lu2021markovian} formulates the circuit design task as a path search problem within a directed graph. In this graph, each node represents a specific gate block, defined as a depth-1 quantum circuit, while the edges indicate the permissible connections between blocks according to predefined rules that prevent redundant operations. MQNE builds quantum circuits block by block. To generate the initial population, the algorithm starts from a fixed node in the directed graph and uniformly samples a certain number of paths of a specific length. In each iteration, the circuits are trained to minimize a predefined loss function. Their fitness is then evaluated based on post-training performance, and the top-performing candidates are selected for the next generation. The algorithm evolves these candidates by appending new gate blocks directly to the end of the selected circuits. This addition of new blocks follows a Markovian process, where the valid choices for the next block depend entirely on the last block of the current circuit. Furthermore, the algorithm constructs circuits using a specific gate set consisting of single-qubit rotation gates and two-qubit controlled $R_x$ gates. Since these gates can act as identity operations when their rotation angles are zero, the circuits in the new generation preserve the capabilities of their parent circuits while exploring new structures. By systematically restricting invalid connections and preserving past performance, MQNE effectively discovers quantum circuits for classifications of real-life images and symmetry-protected topological states.

The quantum circuit evolution of augmenting topologies (QCEAT) algorithm \cite{huang2022robust} encodes quantum circuits as variable-length genomes by representing each quantum gate as a gene that specifies the gate type, the target qubits, and the gate parameters. The algorithm initializes a population and divides it into distinct species based on structural similarity of the circuits, where the distance between two circuits is measured by the minimum number of basic actions needed to convert one genome into the other. The evolutionary process is organized into an inner loop and an outer loop. Within the inner loop, each species evolves independently through mutation operations that randomly add, remove, or substitute quantum gates, and intraspecies crossover operations that generate offspring by aligning two parent circuits to preserve matched components while recombining unmatched segments. These newly generated candidates then undergo a training phase to fine-tune their gate parameters. A selection step subsequently filters these circuits based on their optimized performance to form an evolved species. In the outer loop, an interspecies crossover recombines circuits from different evolved species, followed by gate parameter fine-tuning and a selection step that retains the best circuits for the next generation. Notably, QCEAT incorporates realistic error models, such as coherent noise and incoherent dephasing. Simulations suggest that evaluating fitness under these noise effects naturally penalizes circuits with an excessive number of gates, guiding the algorithm to converge on shallow, noise-resilient circuits.

In evolutionary algorithms, a large number of individuals have to be evaluated to determine their fitness. This process is time-consuming, as the underlying circuit training is prohibitively expensive. To address this computational bottleneck, alternative evaluation methods have been developed to bypass full training (see Section \ref{Sect:efficient} for further elaboration). For instance, performance predictors and training-free proxies have been leveraged in QAS using multi-objective evolutionary algorithms \cite{soloviev2024trainability}. In this approach, an initial population is generated based on a predefined probability distribution, and the ground-truth performance of these circuits is obtained through training. Subsequently, a support vector machine (SVM) is trained to compare the performance of circuit pairs. In each evolutionary cycle, a training-free proxy, referred to as ``information content'' \cite{perez-salinas2024analyzing}, is calculated for each circuit to reflect its trainability. Each circuit is also assigned a score based on pairwise comparisons performed by the SVM. By combining the information content and the score, the fitness of each individual is determined. The top individuals are selected and trained to obtain their actual performance. Both the SVM and the sampling distribution are updated using these top individuals. Finally, the next generation of circuits is sampled according to the updated distribution.

In summary, existing studies suggest that the population-based search mechanism of evolutionary algorithms can help maintain structural diversity during the QAS process and reduce the risk of premature convergence to suboptimal circuit structures. The key to leveraging these algorithms effectively lies in the representation of circuit structures and the design of evolutionary operators. 
The choice of circuit encoding can strongly influence the effectiveness of crossover and mutation operators. Well-crafted evolutionary operators can incorporate domain-specific knowledge to ensure that new generations inherit functional advantages while exploring potentially superior structures. Conversely, poorly designed representations or operators may lead to inefficient sampling of the vast search space and slow convergence. In practical QAS, hardware constraints such as native gate sets, qubit connectivity, and noise characteristics are increasingly taken into account.

\subsection{Bayesian Optimization}
\label{Sect:BO}

Bayesian optimization is a method for optimizing an expensive-to-evaluate function when each evaluation requires significant time or resources. The process starts by evaluating the function at a small number of initial points. Then a probabilistic model, typically a Gaussian process, is built to approximate the function and provide an estimate of uncertainty. Based on this model, an acquisition function is defined to guide the selection of the next evaluation point. This function balances exploration, which favors regions with high uncertainty, and exploitation, which focuses on regions expected to have good values. The algorithm iteratively selects new points by optimizing the acquisition function, evaluates the true function at those points, and updates the probabilistic model. By repeating this cycle, the method efficiently approaches the optimum with relatively few evaluations. Bayesian optimization is widely used in hyperparameter tuning in machine learning, as well as in engineering design, experimental science, and other settings where evaluations are expensive.

A quantum circuit metric was developed to facilitate Bayesian optimization using Gaussian processes \cite{duong2022quantum}. Specifically, a geometrically meaningful distance between quantum gates was proposed and then extended to the circuit level through optimal transport. This distance serves as the basis for constructing a custom kernel function for the Gaussian process. The acquisition function is optimized using an evolutionary strategy, where new candidate circuits are produced through random mutations of gate types and target qubits. Circuit structures yielding higher acquisition values are prioritized for survival. This QAS approach was evaluated across three distinct tasks: quantum compiling, QAOA, and a quantum generative adversarial network consisting of a quantum generator and a classical discriminator \cite{situ2020quantum}.

A graph neural network-based performance predictor has been incorporated into the Bayesian QAS framework \cite{he2025self}. The predictor takes as input a directed acyclic graph (DAG) that represents the structure of a quantum circuit, where nodes correspond to quantum gates and edges encode the dependency relationships between gates. The QAS process starts by sampling a small set of quantum circuits and obtaining their ground-truth performance via circuit training. These labeled circuits are used to train an initial predictor. In each iteration, acquisition scores for circuits in the unlabeled circuit pool are estimated using an ensemble of predictors. The circuits with the highest acquisition scores are then selected for ground-truth performance evaluation and added to the labeled set. The predictor is subsequently updated using all available labeled circuits. This iterative process continues until the predefined evaluation budget is reached. The algorithm ultimately returns the quantum circuit that achieves the best observed ground-truth performance. The key feature of this work lies in the self-supervised learning of the performance predictor through expressibility prediction. Expressibility is an intrinsic metric that characterizes a variational quantum circuit's ability to uniformly explore the Hilbert space as its parameters are varied \cite{sim2019expressibility}. This pretraining strategy enables the predictor to learn representations that are closely aligned with the functional properties of quantum circuits, resulting in a structured and smooth latent space in which circuits with similar functional capabilities are naturally clustered together. Such a representation facilitates more effective navigation of the search space by Bayesian optimization, thereby improving its ability to identify high-potential circuit structures.

While high expressibility is essential for variational quantum circuits to encompass the desired solution, it often comes at the expense of trainability. Highly expressive circuit structures are prone to flat cost landscapes, which significantly hinders gradient-based training \cite{holmes2022connecting}. A recent study considered how to identify expressible and trainable variational quantum circuits by leveraging Bayesian optimization \cite{roseler2026how}. In that work, a practical trainability objective is defined through empirical gradient statistics, utilizing a derived finite-sample and dimension-independent concentration bound for cost function variance estimation. The Bayesian QAS framework targets the joint optimization of circuit properties including expressibility, trainability, entanglement, and circuit complexity (defined by the number of parameters, depth, and gate count). In a quantum neural network application, the search identifies circuit structures with improved effective dimension using over six times fewer parameters compared to a reference structure. For VQE applied to the hydrogen molecule, the framework achieves accuracy comparable to the unitary coupled-cluster singles and doubles ansatz at substantially reduced circuit complexity.

To summarize, existing studies suggest that Bayesian optimization can effectively balance exploration and exploitation in the search over quantum circuit structures, particularly under limited evaluation budgets. The integration of domain-specific representations, such as quantum circuit metrics or learned embeddings from graph neural networks, allows the underlying model to discover the functional similarities between different circuit structures, leading to more informed acquisition decisions. From a methodological perspective, however, we note that Bayesian optimization may face scalability limitations. In particular, Gaussian process-based Bayesian optimization becomes increasingly expensive as the number of evaluated circuits grows and may face difficulties in  high-dimensional search spaces.

\subsection{Reinforcement Learning}
\label{Sect:RL}

Reinforcement learning is a paradigm within machine learning concerned with optimizing sequences of decisions. In this setting, an agent engages with an environment in an iterative manner. At each step, it receives an observation that characterizes the current state, and then selects an action accordingly. The environment responds by transitioning to a new state and providing a reward that reflects the effectiveness of the chosen action. Through this interaction process, the agent aims to learn a policy, which specifies how actions should be selected in different states so as to maximize the expected cumulative reward. Reinforcement learning is broadly applicable across domains such as robotic control, autonomous driving, and game intelligence.

Deep reinforcement learning has been utilized to compile an arbitrary single-qubit gate $U$ into a sequence of elementary gates \cite{zhang2020topological}. The task is formulated as finding a gate sequence $g_n \dots g_1 \approx U$, which, by applying inverse operations, is equivalent to finding a path from $U$ back to the identity matrix $I$. The minimum cost, defined as the number of gates required to reach the identity matrix from a unitary matrix $s$, is expressed by the cost-to-go function $J(s)$:
\begin{align}
J(s) = \min_a \left(g(s, a) + J(S(s, a))\right).
\end{align}
In this equation, $g(s, a)$ represents the cost of taking action $a$, and $S(s, a)$ is the new unitary matrix resulting from that action. A neural network is trained to approximate this function using data generated by applying random gate sequences to the identity matrix. The network then serves as a heuristic function to guide a modified weighted A\texttt{*} algorithm. This algorithm generates gate sequences using the evaluation function:
\begin{align}
f(s) = \lambda G(s) + J(s) + d(s).
\end{align}
Here, $G(s)$ represents the actual cost from the initial state to the current state $s$. The parameter $\lambda \in [0, 1]$ enables a trade-off between search speed and solution optimality. A smaller $\lambda$ causes the algorithm to select the next state to expand primarily based on the heuristic estimate $J(s)$, thereby reducing the number of explored states but may result in longer gate sequences. Additionally, $d(s)$ is a decimal penalty term designed to prioritize training states with near-integer $J(s)$ values over those with decimal parts, as the latter often stem from less reliable estimations or interpolations. When applied to the topological compiling of Fibonacci anyons, this method produces near-optimal braiding sequences for arbitrary single-qubit unitaries.

In variational quantum compiling, reinforcement learning is employed to construct a circuit from a native gate set that becomes functionally equivalent to a target unitary \cite{he2021variational}. An agent is trained to sequentially select gates and their operated qubits using double Q-learning with an $\epsilon$-greedy exploration strategy and experience replay. Within this approach, the agent initially explores the search space by generating random quantum circuits and then transitions toward exploitation to identify higher-performing circuits based on its findings. Simulation results for two-qubit and three-qubit unitaries demonstrate that this approach achieves exact compilation with fewer gates than previous variational quantum compiling algorithms.

In addition to quantum compiling, a deep reinforcement learning-based QAS method was introduced to design quantum circuits for preparing target quantum states \cite{kuo2021quantum}. In this method, an agent interacts with a quantum simulator or physical quantum hardware. At each time step, the agent selects an action from a predefined set of elementary gates. The chosen action is applied to update the current circuit, after which the environment executes the circuit and evaluates its fidelity with respect to the target state. If the fidelity exceeds a predefined threshold, the process terminates and the agent receives a large positive reward. Otherwise, a small negative reward is assigned to encourage the discovery of shorter gate sequences. The observations returned by the environment consist of the expectation values of Pauli $X$, $Y$, and $Z$ operators on each qubit. The interaction continues until the fidelity threshold is reached or the maximum number of steps is exceeded. Reinforcement learning algorithms such as advantage actor-critic (A2C) and proximal policy optimization (PPO) are employed to optimize the agent. Simulation results demonstrate that the agent can successfully generate quantum gate sequences for preparing Bell states and GHZ states, and that PPO generally exhibits more stable and faster convergence than the A2C method.

A feedback-driven curriculum learning technique has been incorporated into reinforcement learning-based QAS, in which the difficulty of the task is dynamically adjusted according to the agent's current performance \cite{ostaszewski2021reinforcement}. In this approach, reinforcement learning is used to discover VQE ansatzes that yield accurate ground state energy estimates. The state representation encodes the current circuit structure, augmented with the corresponding energy estimate. The action space consists of quantum gates that can be appended to the circuit. A double deep-Q network with an $\epsilon$-greedy policy is employed to guide the learning process. Task difficulty is characterized by an energy error threshold that the agent must satisfy in order to receive a reward. Rather than requiring the agent to achieve chemical accuracy from the outset, the method begins with a more attainable target. As the agent identifies circuits that meet the current requirement, the threshold is progressively reduced based on the best results obtained so far. To mitigate the risk of stagnation or failure when the task becomes overly challenging, an amortization mechanism is introduced, which temporarily relaxes the threshold and provides additional flexibility for continued learning. Building on the aforementioned approach, curriculum-based reinforcement learning quantum architecture search (CRLQAS) \cite{patel2024curriculum} further improves performance and stability in noisy quantum environments. It introduces a tensor-based binary encoding of quantum circuits, an illegal action mechanism to prune the search space, and a random halting scheme to encourage compact circuits. In addition, it employs a tailored Adam-SPSA optimizer for enhanced convergence speed and robustness, together with a noisy quantum circuit simulator based on the Pauli-transfer matrix formalism.

To provide the agent with access to the internal multipartite entanglement structure of quantum states, an entanglement-aware QAS method based on deep reinforcement learning was proposed for state synthesis tasks \cite{bi2025general}. This method enriches the reward signal by integrating fidelity with a quantitative measure of multipartite entanglement. Specifically, it incorporates the absolute difference between the entanglement of the target state and that of the current state. As a result, the agent is endowed with a more comprehensive physical characterization of the state space, allowing it to distinguish between states that are similar in fidelity yet structurally different. Simulation results demonstrate that the entanglement-aware agent outperforms its fidelity-driven counterpart.

Gadget reinforcement learning (GRL) was introduced into QAS \cite{kundu2026reinforcement}. The framework first employs a reinforcement learning agent to solve simple problem instances using a basic action space composed of native gates. The best-performing circuits are then analyzed through a program synthesis module, which identifies frequently recurring gate patterns and encapsulates them into reusable composite operations, referred to as gadgets. These gadgets are subsequently incorporated into the action space, allowing the agent to operate with higher-level building blocks instead of relying solely on native gates. Equipped with this enriched action space, the agent is retrained to address more complex problem instances more efficiently.

In conclusion, the studies reviewed in this subsection indicate that reinforcement learning-based QAS provides a flexible approach to exploring quantum circuit structures by formulating the design process as a sequential decision-making problem. The reported results also highlight the importance of reward function design, as poorly specified rewards may lead to unstable training or suboptimal solutions. Techniques such as curriculum learning and the incorporation of entanglement-based metrics have been used to construct more informative reward signals. The choice of reinforcement learning algorithm can likewise influence the balance between exploring the large combinatorial search space and exploiting high-performing structures. Furthermore, we identify the potential for transferable search strategies as an additional advantage of reinforcement learning-based QAS, where a learned policy from previous tasks may be transferred to accelerate the exploration of new circuit structures. However, we consider sample inefficiency an important limitation of reinforcement learning-based QAS, since extensive interactions with simulators or hardware can incur substantial computational cost, particularly for large-scale quantum circuits.

\subsection{Monte Carlo Tree Search}
\label{Sect:MCTS}

Monte Carlo tree search is a decision-making method that explores possible actions by evaluating many future outcomes. It builds a tree of states and actions, repeatedly selecting promising paths, expanding new nodes, performing randomized rollouts, and updating the results to favor better choices. By balancing exploration and exploitation, it can identify good decisions without exhaustively searching the entire space. It is widely used in game playing and planning.

A QAS algorithm based on Monte Carlo tree search has been proposed \cite{meng2021quantum}. In this algorithm, the search space is modeled as a Monte Carlo tree, which can be regarded as a supernet encompassing all possible circuit structures. Each circuit corresponds to a path in the tree, where the $i$-th layer of the circuit is represented by a node at the $i$-th level of the tree. To eliminate redundant gates, a set of predefined rules is imposed to constrain the candidate nodes. To reduce computational overhead, a weight sharing (see Section \ref{Sect:weight} for further elaboration) strategy is employed across different circuit structures within the supernet. During the supernet warm-up phase, random paths are iteratively sampled from the tree to instantiate circuit structures, and the corresponding circuit parameters undergo a single update using gradient-based optimization. A similar procedure is followed in the subsequent supernet training phase, where random path sampling and single-step parameter updates are repeatedly performed. Throughout training, each node in the tree maintains two statistics: a Q-value, which reflects the quality of the node (measured via the loss), and a visit count, which records how frequently the node has been selected. For candidate circuit generation, a hierarchical node selection strategy based on the upper confidence bounds for trees criterion is adopted. This mechanism leverages the stored node statistics to balance exploration and exploitation effectively during sampling. Finally, the sampled candidate circuits are fully trained to obtain their true performance, and the circuit achieving the best performance is selected as the final solution. 

Another study introduced a combinatorial multi-armed bandit model, together with a naive assumption, to characterize the selection of unitary operators at each circuit layer \cite{wang2023automated}. Under this framework, the reward of a fully constructed circuit is used to linearly approximate the rewards of different unitaries, and the search process is carried out using a nested Monte Carlo tree search algorithm. The resulting approach was validated on several tasks, including VQE, QAOA, variational quantum linear solver \cite{bravo2023variational}, and the design of encoding circuits for quantum error detection codes.

An alternative formulation of the QAS problem within a tree structure was introduced in progressive widening enhanced Monte Carlo tree search (PWMCTS) \cite{lipardi2025quantum}. In this framework, nodes represent quantum circuits and edges correspond to modification actions such as appending, replacing, or deleting a random gate, and tuning a random gate's parameter. A progressive widening technique is adopted to dynamically explore the search space, addressing the challenge of an infinite number of possible actions. The search process is gradient-free, only the quantum circuits identified by PWMCTS are subsequently fine-tuned using gradient-based optimization.

Monte Carlo graph search has been applied to QAS by identifying and merging equivalent circuits that are represented by different nodes reached through distinct paths, thereby reducing redundancy \cite{rosenhahn2023monte}. In this approach, each node represents a quantum circuit, and a directed edge exists between two nodes if one circuit can be transformed into another by adding a single gate. The nodes are decorated with the resulting unitary obtained by concatenating the operations along the shortest path from the start node. The algorithm samples a node based on its task-specific quality score and then selects a gate to expand the circuit. If the unitary matrix of the new circuit is equivalent to one represented by an existing node within a small numerical threshold, an edge is added from the sampled node to the existing node. Otherwise, a new node and edge are created. When a node that solves the task is reached, the final quantum circuit is obtained by identifying the shortest path from the start node to that specific solution.

Overall, existing studies suggest that Monte Carlo tree search-based QAS methods provide a versatile approach to navigating the search space by formulating circuit design as a sequential decision problem. Their tree-based search enables a systematic balance between exploration and exploitation and provides an interpretable record of the search process. Several techniques have been introduced to improve search efficiency and scalability. For example, weight sharing has been employed to reuse parameters across different circuit structures and reduce the cost of repeated training, while progressive widening and graph-based merging have been used to control tree growth and reduce redundant exploration. From a methodological perspective, however, we note that Monte Carlo tree search-based QAS also has several inherent limitations. Its performance can be sensitive to the choice of tree expansion strategy and rollout policy, as well as the accuracy of reward estimation. Moreover, maintaining statistics for an increasing number of tree nodes can introduce substantial memory overhead as the combinatorial search space grows.

We compare the four search strategies reviewed in this section in Table~\ref{tab:discrete_strategies}. Based on the studies reviewed above, these strategies differ not only in their search mechanisms but also in how information obtained during the search is reused to guide subsequent exploration. Evolutionary algorithms maintain and evolve a population of candidate circuits, which promotes structural diversity and broad exploration but generally requires evaluating a relatively large number of candidates. Bayesian optimization instead relies on a surrogate model to guide the selection of promising circuits and is therefore particularly attractive when the circuit evaluation budget is limited, although its effectiveness depends strongly on the surrogate model and acquisition function. Reinforcement learning and Monte Carlo tree search both formulate circuit construction as a sequential decision-making process, but they exploit accumulated search information in different ways: reinforcement learning encodes experience into a learned policy that can potentially be reused across search trajectories or problem instances, whereas Monte Carlo tree search explicitly organizes experience within a search tree and provides a more interpretable record of the explored trajectories. These differences lead to distinct practical trade-offs. Reinforcement learning can offer adaptive and transferable search behavior but often requires substantial interaction data and careful reward and policy design, while Monte Carlo tree search avoids learning an explicit policy but can incur considerable memory overhead and is sensitive to expansion and rollout strategies. Overall, understanding the characteristics of these search strategies is important for selecting suitable search algorithms for specific QAS problems and for identifying potential directions for future improvements.

\begin{table}
\centering
\caption{Comparison of search strategies reviewed in Section \ref{Sect:discrete}}
\label{tab:discrete_strategies}
\begin{tabular}{|c|c|c|c|c|} 
\hline
\multicolumn{1}{|l|}{}                                                    & \begin{tabular}[c]{@{}c@{}}\textbf{Evolutionary}\\\textbf{Algorithm}\end{tabular}                             & \begin{tabular}[c]{@{}c@{}}\textbf{Bayesian}\\\textbf{Optimization}\end{tabular}                             & \begin{tabular}[c]{@{}c@{}}\textbf{Reinforcement}\\\textbf{Learning}\end{tabular}                         & \begin{tabular}[c]{@{}c@{}}\textbf{Monte Carlo}\\\textbf{Tree Search}\end{tabular}                                         \\ 
\hline
\begin{tabular}[c]{@{}c@{}}\textbf{Core}\\\textbf{Mechanism}\end{tabular} & \begin{tabular}[c]{@{}c@{}}population-based\\evolution via\\mutation,\\crossover, and\\selection\end{tabular} & \begin{tabular}[c]{@{}c@{}}surrogate-guided\\search with\\exploration-\\exploitation\\trade-off\end{tabular} & \begin{tabular}[c]{@{}c@{}}sequential\\decision-making\\with\\reward-driven\\policy learning\end{tabular} & \begin{tabular}[c]{@{}c@{}}tree-based search\\using selection,\\expansion,\\simulation, and\\backpropagation\end{tabular}  \\ 
\hline
\textbf{Advantages}                                                       & \begin{tabular}[c]{@{}c@{}}high structural\\diversity, strong\\global exploration\end{tabular}                & \begin{tabular}[c]{@{}c@{}}sample-efficient\\search\end{tabular}                                             & \begin{tabular}[c]{@{}c@{}}adaptive policy\\learning,\\transferable\\search strategies\end{tabular}       & \begin{tabular}[c]{@{}c@{}}balanced\\exploration and\\exploitation,\\interpretable\\search trajectory\end{tabular}         \\ 
\hline
\textbf{Considerations}                                                   & \begin{tabular}[c]{@{}c@{}}individual\\ encoding, design \\ of evolutionary\\operators\end{tabular}                                    & \begin{tabular}[c]{@{}c@{}}choice of\\surrogate model\\ and acquisition\\ function\end{tabular}                                           & \begin{tabular}[c]{@{}c@{}}reward function\\design,\\ policy selection \end{tabular}                                           & \begin{tabular}[c]{@{}c@{}}expansion and\\ rollout strategies,\\high memory\\consumption\end{tabular}                                                           \\
\hline
\end{tabular}
\end{table}

\section{Continuous Relaxation and Representation of Architecture Spaces}
\label{Sect:d2c}

While the approaches discussed in Section \ref{Sect:discrete} search directly over discrete circuit structure spaces, another category of QAS methods formulates the search space in a continuous form to enable optimization in a continuous space. This section reviews two classes of approaches within this category.
The first introduces structure parameters to relax the discrete circuit design space into a differentiable formulation, enabling gradient-based optimization of both circuit structures and gate parameters. The second maps circuit structures into a continuous latent representation using variational autoencoders and performs optimization within the learned latent space.

\subsection{Differentiable Quantum Architecture Search}
\label{Sect:DQAS}

The discrete search space of quantum circuit structures can be relaxed to a continuous and differentiable domain. A differentiable quantum architecture search (DQAS) approach has been proposed to enable end-to-end automated circuit design \cite{zhang2022differentiable}. In this approach, a circuit is represented as a sequence of $p$ primitive unitaries selected from an operation pool that includes one-qubit gates, two-qubit gates or higher-level building blocks. One operation from the pool can be used multiple times in building a single circuit. The sampling distribution is modeled independently for each position in the gate sequence and governed by a structure parameter $\alpha\in \mathbb{R}^{p\times c}$, where $c$ is the number of candidate operations in the pool. Specifically, $\alpha_{ij}$ denotes the logit associated with placing the $j$-th operator from the pool at the $i$-th position, so the probability of selecting the $j$-th operator at the $i$-th position is given by a softmax $e^{\alpha_{ij}}/\sum_k e^{\alpha_{ik}}$. The structure parameter $\alpha$ is initialized to zero, which results in a uniform sampling distribution at the beginning of training. 
The gate parameters $\theta$ are organized as a tensor of size $p\times c\times l$, where $l$ is the maximum number of parameters among candidate operations. When the same operation is selected at the same position across different sampled circuits, the corresponding parameters are shared. The joint optimization of structure and gate parameters proceeds as follows. A batch of circuits is sampled according to the distribution defined by $\alpha$. The loss function associated with the target task is evaluated, and gradients with respect to both $\alpha$ and $\theta$ are computed. These parameters are then updated using gradient-based optimization. This procedure is repeated until convergence. After training, the circuit structure with the highest probability is selected, and its associated gate parameters are further fine-tuned. Although the sampling step is not differentiable, gradients with respect to $\alpha$ can be estimated using the score function estimator.

Another differentiable QAS method, QuantumDARTS, has also been proposed \cite{wu2023quantumdarts}. This method constructs an $n$-qubit circuit with $m$ layers by sampling from a set of structure parameters $\alpha$. For the gate applied to the $i$-th qubit at the $j$-th layer, the probability of selecting the $k$-th candidate gate is given by $\exp(\alpha_{ij}^{(k)})/\sum_{k’} \exp(\alpha_{ij}^{(k')})$. To enhance efficiency, the Gumbel-Softmax technique is employed for circuit sampling. Specifically, it samples every gate in the circuit by selecting the candidate with the maximal  structure parameter augmented by Gumbel noise during the forward pass, while substituting this discrete choice with a Softmax-based relaxation during the backward process to enable gradient back-propagation for updating structure parameters. The framework provides two variants. The macro search version targets full circuit optimization, while the micro search version focuses on discovering transferable subcircuit structures that can be applied to larger-scale problems.

A meta quantum architecture search (MetaQAS) method was proposed to leverage meta-learning for capturing cross-task knowledge \cite{he2022quantum}. The goal is to learn effective initialization strategies for both structure parameters and gate parameters in differentiable QAS. The meta-learning algorithm Reptile is adopted. The meta-parameters for structure and gate parameters are randomly initialized. During each training iteration, a mini-batch of tasks is sampled from the task distribution. For each task, the QAS procedure is initialized using the current meta-parameters, and the structure parameters and gate parameters are updated for $k$ gradient steps. After task-specific adaptation, the meta-parameters are updated by moving them toward the task-adapted parameters as follows:
\begin{align}
& \theta_{meta}^{j+1}=\theta_{meta}^{j} + \lambda_{meta}\frac{1}{m}\sum_{i=1}^{m}(\theta_{\mathcal{T}_i}^{k}-\theta_{meta}^{j})\\
& \alpha_{meta}^{j+1}=\alpha_{meta}^{j} + \eta_{meta}\frac{1}{m}\sum_{i=1}^{m}(\alpha_{\mathcal{T}_i}^{k}-\alpha_{meta}^{j})
\end{align}
Here, $\theta_{\mathcal{T}_i}^{k}$ and $\alpha_{\mathcal{T}_i}^{k}$ denote the gate parameters and structure parameters obtained after $k$ update steps on task $i$. The coefficients $\lambda_{meta}$ and $\eta_{meta} $ are the learning rates for updating the meta-parameters. This process is repeated until convergence. After convergence, the learned meta-parameters provide a favorable initialization for new tasks. This initialization enables the QAS algorithm to adapt to unseen tasks with fewer gradient updates, thereby improving learning efficiency and accelerating convergence.

In summary, existing studies suggest that differentiable QAS introduces structure parameters to represent discrete structural choices probabilistically. This formulation replaces explicit combinatorial search over circuit structures with gradient-based optimization in a continuously relaxed search space, thereby enabling the joint optimization of structure and gate parameters within a unified framework. From a methodological perspective, however, we note that this efficiency comes at the cost of relying on a continuous surrogate for an inherently discrete circuit structure space. In particular, independently parameterizing operation choices at different circuit positions may limit the ability to capture dependencies among gates. In addition, the circuit obtained from the relaxed optimum after structure selection may not correspond to the best circuit in the original discrete search space. These considerations suggest that the effectiveness of differentiable QAS depends not only on optimization efficiency but also on how faithfully the continuous relaxation preserves the structural relationships and relative performance of the circuits in the original discrete search space.

\subsection{Latent Representation-based Quantum Architecture Search}
\label{Sect:latentQAS}

Differentiable QAS relaxes the discrete search space into a continuous domain via a parameterized probabilistic model. To mitigate the resource intensive sampling overhead of differentiable QAS, gradient-based optimization for quantum architecture search (GQAS) \cite{he2024gradient} employs a variational graph autoencoder to learn continuous latent representations of circuit structures, and then optimizes these representations to maximize the output of a performance predictor. In the pre-training phase, the encoder and decoder are trained through a reconstruction task. The encoder maps the DAG of a circuit structure to a latent vector, while the decoder reconstructs the circuit structure from this vector. A candidate set is built by randomly sampling quantum circuits from the search space, and these circuits are trained to obtain their ground truth performances. In the model training phase, a multilayer perceptron predictor is trained on the candidate set to learn the mapping between latent vectors and circuit performances. When the predictor is trained, the encoder and decoder are fine-tuned in an end-to-end manner by jointly minimizing the performance prediction loss and the reconstruction loss. In the structure optimization phase, the top performing $K$ circuits in the candidate set are selected and mapped to latent vectors. These latent vectors are then fed into the predictor and optimized through gradient-based methods to maximize the predicted performance. The optimized latent vectors are subsequently decoded into circuit structures. These newly generated circuits are trained to obtain their ground-truth performances and are added to the candidate set. The model training phase and the structure optimization phase are then repeated iteratively. Finally, the circuit with the best performance in the candidate set is selected as the output.

Another QAS framework also employs a variational autoencoder to learn continuous latent representations of circuit structures \cite{sun2026quantum}. Instead of relying on a predictor to guide the optimization, it adopts the reinforcement learning algorithm REINFORCE to explore the latent space. The environment state is represented by the latent vector, and the agent is implemented using a recurrent neural network. At each step, the agent selects an action that produces a latent vector conditioned on the current state, and this latent vector is then decoded to propose candidate circuit structures for evaluation. In addition, Bayesian optimization is used as an alternative approach to explore the latent space.

To summarize, the studies reviewed above show that mapping discrete circuit structures into a continuous latent space provides an alternative way to perform QAS. Unlike differentiable QAS, which directly relaxes discrete structural choices through continuous structure parameters, latent representation-based methods leverage generative models such as variational autoencoders to learn a smooth and compact representation of circuit structures and then perform optimization within this learned space. This formulation is also compatible with different optimization strategies, including gradient-based optimization, reinforcement learning, and Bayesian optimization. Such an approach can lead to improved sample efficiency and better scalability, as optimization is carried out in a lower-dimensional, structured latent space rather than directly over discrete circuit structures. However, we note that the effectiveness of these methods appears to depend strongly on the quality of the learned latent space. In particular, the learned space should preserve meaningful structural relationships among the original circuits, while the decoder should reliably map optimized latent vectors back to valid and high-performing circuit structures. Otherwise, optimization in the latent space may not translate into corresponding improvements in the original discrete circuit structure space.

\section{Efficient Performance Estimation}
\label{Sect:efficient}

Efficient performance estimation plays a crucial role in QAS, as directly evaluating circuit performance through full training is expensive on both simulators and quantum hardware. To address this challenge, a variety of estimation methods have been developed to approximate circuit quality with significantly reduced cost. These methods aim to provide reliable performance signals that can guide the search process while avoiding repeated optimization of candidate circuits. In this section, we review three classes of approaches, namely weight sharing mechanism, predictor-based methods, and training-free proxies.

\subsection{Weight Sharing}
\label{Sect:weight}

Weight sharing has emerged as an efficient estimation method for reducing the parameter search space. Instead of training each candidate circuit from scratch, the weight sharing mechanism enables rapid performance estimation by reusing parameters across different circuit structures according to specified rules.

A one-stage QAS method was proposed, which trains a supernet only once \cite{du2022quantum}. It defines a search pool of candidate circuit structures based on the number of qubits, a maximum circuit depth, and a predefined set of allowed gate types. Each circuit consists of multiple layers, and each layer is composed of a sequence of parameterized single-qubit and two-qubit gates. To efficiently represent this large pool, the method constructs a supernet with two roles. First, it serves as an indicator of the search space through an indexing technique, where each candidate structure corresponds to an index list that specifies the gate choices at each layer. Second, it parameterizes all candidate structures through a weight sharing strategy that correlates parameters across different circuits. Specifically, when the layout of single-qubit gates at a given layer is identical across different circuit structures, the corresponding parameters are shared. During training, a circuit structure is uniformly sampled from the pool at each iteration, and the parameters associated with the sampled structure undergo a single update. After sufficient training, a set of $K$ circuits is sampled uniformly and ranked, and the best performing circuit is selected for further parameter fine-tuning. To address the difficulty of accurately optimizing correlated parameters shared among all candidate structures, the method employs multiple supernets instead of a single one. Each supernet is initialized and trained independently. During both training and ranking, each supernet applies weight sharing to parameterize the sampled circuit, and the parameters used for that circuit are selected from the supernet that achieves the lowest loss.

A framework called noise-adaptive search for robust quantum circuits (QuantumNAS) \cite{wang2022quantumnas} was proposed for noise-resilient co-search of circuit structures and qubit mappings. It first constructs a supercircuit by stacking a sufficient number of layers of predefined parameterized gates to cover a large design space. The supercircuit is trained by repeatedly sampling subcircuits and updating their parameters. Each subcircuit corresponds to a sampled subset of gates from the supercircuit, and its parameters are inherited through weight sharing. After training, an evolutionary co-search is conducted to jointly identify the most robust subcircuit and qubit mapping. During this process, the performance of each subcircuit is estimated using inherited parameters and simulated under realistic device noise models. Once the best circuit and mapping pair is identified, the circuit is retrained from scratch. Finally, iterative pruning is applied to reduce noise. Redundant gates with small parameter magnitudes are removed, followed by fine-tuning to recover performance. The pruning ratio is gradually increased with repeated fine-tuning until the target sparsity level is achieved. 

In conclusion, the works discussed in this subsection indicate that weight sharing can substantially reduce the cost of candidate evaluation by allowing different circuit structures to reuse parameters rather than training each candidate independently. Weight sharing therefore serves as a core mechanism in differentiable QAS and is tightly integrated with the search process. At the same time, the reported results also reveal a fundamental limitation: parameter sharing couples the optimization of different circuit structures, so the inherited parameters may not accurately reflect the standalone performance of individual candidates. This can lead to inconsistencies between performance estimates obtained during the search and those obtained after independent training.

\subsection{Performance Predictors}
\label{Sect:predictor}

In the predictor-based QAS approach, a neural network predictor is used to directly estimate the performance of quantum circuits based solely on their structures \cite{zhang2021neural}. During the predictor training phase, random circuits are generated in a gate-wise or layer-wise manner. These circuits are trained, and a dataset consisting of circuit structures and their corresponding performance is obtained. The predictor is then trained on this dataset. It takes an image representation of the circuit structure as input. During the screening phase, the trained predictor is used to filter a large number of randomly generated circuit candidates. Only a small subset with predicted performance above a given threshold is selected for further evaluation through circuit training. The circuit with the best performance is chosen as the final result. To address the long tail distribution problem, which is difficult to model with a single regression model, a two-stage screening strategy is adopted. First, a convolutional neural network-based binary classifier is trained to determine whether a candidate is promising. Only the promising circuits are passed to a recurrent neural network-based regression model for more fine-grained evaluation. In addition, a beam search-based method is developed to transfer optimal circuit structures identified on smaller systems to larger systems.

In predictor-based QAS, the predictor is trained through supervised learning. The main challenge lies in learning an accurate predictor from a limited number of labeled circuits. To address this issue, graph self-supervised quantum architecture search (GSQAS) \cite{he2023gsqas} introduced a self-supervised learning strategy that enables the training of a precise predictor with significantly fewer labeled samples. This approach first pretrains a graph encoder using a carefully designed pretext task on a large set of unlabeled quantum circuits, with the goal of learning informative representations of circuit structures. In the pretext task, the encoder takes the DAG representation of a circuit structure and embeds it into a continuous latent representation, while a decoder reconstructs the original circuit structure from this representation. Since this stage relies only on unlabeled circuits, it is independent of any specific task. Once pretrained, the encoder can be transferred to downstream tasks. In this stage, the encoder output is fed into a multilayer perceptron predictor. The encoder and the predictor are then fine-tuned jointly using a small set of labeled circuits associated with a specific VQA task. 

In addition to self-supervised learning, semi-supervised learning has been introduced in two predictor-based QAS methods to leverage the latent information contained in unlabeled circuits \cite{situ2026data}. The first method follows a teacher-student framework, in which a teacher model guides the training of a student predictor using unlabeled data. The teacher generates smoother and more stable targets, which encourage the student to produce consistent predictions under perturbations and improve robustness. The second method adopts an uncertainty-aware circuit selection strategy that identifies unlabeled circuits with low prediction uncertainty. These selected circuits are then added to the training set, which effectively enlarges the labeled dataset.

Overall, predictor-based QAS methods decouple circuit evaluation from circuit training by using a learned performance model to estimate the quality of candidate circuit structures. As demonstrated in existing studies, the low inference cost of such predictors enables a large number of candidate circuits to be evaluated efficiently, and even relatively simple sampling strategies can be effective when combined with an accurate predictor. These methods are also flexible in accommodating different circuit encodings and deep learning models. However, this efficiency introduces a trade-off between estimation cost and estimation fidelity. Inaccurate predictions may misguide the search process and prematurely discard promising circuit structures, making predictor reliability an important practical consideration. Limited labeled data can further degrade prediction accuracy and generalization, motivating the use of self-supervised and semi-supervised learning techniques. Recent studies have also begun to incorporate uncertainty-aware estimation, for example through ensembles or Bayesian predictors, so that prediction confidence can be considered together with estimated performance when selecting candidate circuits \cite{he2025self,situ2026data}. Such uncertainty information can reduce the risk of making structural decisions based on unreliable predictions. Taken together, these studies suggest that improving not only prediction accuracy but also reliability and uncertainty estimation is important for the effective use of performance predictors in QAS.

\subsection{Training-free Proxies}
\label{Sect:training-free}

A training-free quantum architecture search (TF-QAS) \cite{he2024training-free} algorithm has been proposed, which uses two proxies to rank circuit structures. By considering both accuracy and computational cost, a two-stage progressive strategy is adopted. In the first stage, a large number of circuit structures are randomly sampled. A path-based proxy, defined by the number of paths in the DAG of the circuit structure, is used to filter out a substantial portion of unpromising candidates. This proxy can be computed with very low overhead and provides a coarse estimate of the circuit’s topological complexity, which relates to properties such as expressibility and entangling capability. It therefore serves as an efficient preliminary filter to eliminate poor candidates. In the second stage, an expressibility-based proxy with strong rank correlation to circuit performance is applied to select the top-$K$ circuit structures from the remaining candidates. For a specific VQA, the final circuit is chosen from these top candidates based on their actual performance on the target task.

A noise-guided QAS framework was proposed to design performant and noise-robust circuits for classical data classification \cite{anagolum2024elivagar}. It decouples the evaluation of noise robustness and performance by introducing two proxies, namely Clifford noise resilience and representational capacity, enabling early rejection of low-fidelity circuits and reducing circuit evaluation costs. Candidate circuits are sampled together with qubit mappings under hardware connectivity constraints. The sampling process is guided by hardware noise information, including qubit coherence times, single-qubit and two-qubit gate fidelities, and readout fidelities. Parametric gates for data embedding are also selected, and each gate is randomly assigned to encode one dimension of the input data. The framework then computes the Clifford noise resilience score for each circuit to estimate its robustness to noise. Candidate circuits are ranked according to this score, and low fidelity circuits are filtered out. For the remaining candidates, their performance on the target classification task is predicted using representational capacity. A composite score that combines representational capacity and Clifford noise resilience is computed, and the circuit with the highest score is selected as the final result. 

A scalable training-free QAS method was introduced based on landscape fluctuation analysis of the cost function \cite{zhu2025scalable}. The landscape fluctuation is defined as the standard deviation of the approximately normalized cost function. It can diagnose several key properties of variational quantum circuits, including barren plateaus, bad local minima, insufficient expressibility, and overparameterization, enabling accurate prediction of circuit learnability without costly training. This proxy is combined with a streamlined two-level search strategy. In the layer-wise search phase, the algorithm starts from an initial circuit and iteratively adds layers selected from a predefined pool. The construction of the layer pool is formulated as a maximum cardinality matching problem in order to exploit task-specific structure in VQE settings. At each step, the best layer configuration is selected based on the landscape fluctuation and the circuit is updated accordingly. In the subsequent gate-wise redundancy elimination phase, gates are iteratively removed from the circuit. After each removal, the circuit is evaluated using the landscape fluctuation, and gates with minimal impact on performance are pruned.

Although training-free QAS has shown strong potential for automating quantum circuit design for VQAs with minimal computational cost, different VQA tasks and search spaces often favor different training-free proxies. In addition, circuit performance is influenced by multiple factors, which makes it difficult for a single proxy to provide an accurate assessment. To address these limitations, an adaptive QAS framework based on a mixture of experts was proposed to dynamically integrate multiple training-free proxies \cite{he2025adaptive}. This approach combines their strengths and reduces individual biases. Specifically, an expert model is trained for each proxy to directly predict its value, which significantly reduces the cost of evaluating multiple proxies for a large number of circuits. Since these expert models are task-agnostic, they can be reused across different VQAs. A gating network takes the circuit structure as input and outputs weights for each expert. The gating network is trained using a small set of labeled data. 

In summary, the works discussed in this subsection show that training-free QAS estimates circuit quality using proxy metrics without fully training each candidate, resulting in very low evaluation cost. The reported results also indicate that hardware information and task-specific knowledge can be incorporated into proxy design. However, the effectiveness of individual proxies can vary across tasks and search spaces, since circuit performance is determined by multiple factors that may not be fully captured by a single metric. We identify proxy selection and combination as a central challenge for training-free QAS. Adaptive fusion strategies partly address this issue by integrating multiple proxies into a weighted evaluation of circuit quality, although their effectiveness still depends on the quality and complementarity of the selected proxies.

We compare the three efficient performance estimation methods in Table \ref{tab:efficient_estimation}. Based on the studies reviewed above, these methods reduce candidate evaluation cost through different mechanisms: weight sharing reuses parameters across circuit structures, performance predictors learn structure-performance relationships from data, and training-free proxies estimate circuit quality using low-cost indicators. From our comparison, these differences lead to distinct trade-offs. Weight sharing amortizes training cost across candidates but may suffer from inaccurate estimation due to parameter coupling. Performance predictors can leverage deep learning models but depend on their generalization ability. Training-free proxies can incorporate task-specific knowledge but may not generalize well across different tasks and search spaces. From the perspective of the overall QAS process, these performance estimation methods primarily affect how candidate circuit structures are evaluated rather than how the structures are searched, and may therefore be integrated with different search strategies, when the corresponding mechanisms are compatible. Their practical effectiveness, however, ultimately depends on whether they provide sufficiently reliable performance estimates for ranking candidate structures.

\begin{table}
\centering
\caption{Comparison of efficient performance estimation methods}
\label{tab:efficient_estimation}
\begin{tabular}{|c|c|c|c|} 
\hline
\multicolumn{1}{|l|}{} & \textbf{Weight Sharing}                                                              & \textbf{Performance Predictors}                                                                                  & \textbf{Training-free Proxies}                                                                     \\ 
\hline
\textbf{Core Idea}     & \begin{tabular}[c]{@{}c@{}}reuse parameters\\across circuit\\structures\end{tabular} & \begin{tabular}[c]{@{}c@{}}train a surrogate model\\to predict performance\\from structure features\end{tabular} & \begin{tabular}[c]{@{}c@{}}estimates circuit\\performance using\\low-cost indicators\end{tabular}  \\ 
\hline
\textbf{Advantages}    & \begin{tabular}[c]{@{}c@{}} amortized training\\ cost across \\ candidates\end{tabular}            & \begin{tabular}[c]{@{}c@{}} deep learning of\\ structure-performance\\ relationships\end{tabular}                                & \begin{tabular}[c]{@{}c@{}}  incorporate task-\\ specific knowledge \\ into proxy design\end{tabular}                          \\ 
\hline
\textbf{Limitations}   & \begin{tabular}[c]{@{}c@{}}inaccurate due to\\parameter coupling\end{tabular}        & \begin{tabular}[c]{@{}c@{}}generalization ability\\of the predictor\end{tabular}                                 & \begin{tabular}[c]{@{}c@{}}task-dependent\\proxy metrics\end{tabular}                              \\
\hline
\end{tabular}
\end{table}

\section{Applications}
\label{Sect:application}

The most widely used benchmark tasks for QAS research include VQE, QAOA, and classical data classification. QAS has also been applied to a broader range of problems. This section reviews some representative applications of QAS, highlighting its flexibility and potential in diverse VQA tasks.

QAS serves as a powerful tool for quantum compiling, notably through the quantum-assisted quantum compiling (QAQC) approach \cite{khatri2019quantum}. This approach focuses on approximating a target unitary using a circuit with a variable structure. The search process is guided by a cost function derived from the Hilbert-Schmidt test or its local counterpart, which quantifies the distance or overlap between the target and the trial unitaries. By evaluating this cost directly on quantum processors, the approach produces optimized hardware-efficient circuits that strictly respect device-specific constraints, such as restricted qubit connectivity and native gate alphabets.

The kernel method is a powerful class of machine learning techniques that uses a feature map to project data from a low-dimensional input space into a high-dimensional feature space, where complex data distributions become more separable. Instead of computing the mapping explicitly, a kernel function efficiently evaluates inner products between pairs of data points in this feature space to serve as a measure of similarity, allowing algorithms such as SVMs to effectively perform complex non-linear classification tasks.  
Quantum kernel methods \cite{havlicek2019superivsed,schuld2019quantum} extend this method into the quantum domain by utilizing quantum circuits as feature maps to encode classical data $x$ into quantum states $|\psi(x)\rangle$ within an exponentially large Hilbert space. In this context, the structural design of the quantum circuit is critical. 
Earlier research utilized QAS to design quantum kernel circuits by maximizing classification accuracy while simultaneously minimizing circuit complexity \cite{altares2021automatic}. Another approach employed kernel-target alignment, which serves as a reliable surrogate for classification accuracy and can be effectively calculated, as an optimization metric \cite{lei2024neural}. Furthermore, a subsequent advancement explored a strategy that concurrently maximize the probability of successful trials \cite{tannu2019not} and kernel-target alignment to achieve more robust and efficient quantum kernel designs \cite{meng2026hardware}.

Quantum error detection and correction protects quantum information by encoding logical qubits into multiple physical qubits through an encoding circuit, allowing errors on the physical qubits to be detected and corrected. In this context, QAS has been applied to search for encoding circuits for the [[4,2,2]] error detection code, which encodes two logical qubits into four physical qubits and has a code distance of 2 \cite{wang2023automated}. The search objective is defined using the fidelity between the state prepared by a candidate circuit and the target encoded state produced by a reference implementation of the code. The results indicate the potential of QAS to automatically identify device-specific encoding circuits for quantum error detection and correction codes on quantum processors.

A QAS method for thermal state preparation of the Sachdev–Ye–Kitaev model was introduced, which combines reinforcement learning with convolutional neural networks \cite{kundu2025improving}. It adopts an iterative optimization scheme to jointly refine the circuit structure and its parameters. The optimization is driven by a hybrid reward function that incorporates both entropy-related terms and expectation values of the SYK Hamiltonian.

Recent research has leveraged QAS to automate the design of quantum autoencoders \cite{kulshrestha2025neural}. Quantum autoencoders \cite{romero2017quantum} are the quantum counterparts of classical autoencoders, which use neural networks to compress inputs into a lower-dimensional latent space and subsequently reconstruct them, primarily for tasks such as dimensionality reduction, feature extraction, and data denoising. Specifically, quantum autoencoders utilize variational quantum circuits to map high-dimensional quantum states onto a reduced set of latent qubits. This process involves mapping the input into a latent subsystem and an isolated trash subsystem. The optimization is guided by a cost function based on the SWAP test, which measures the fidelity between the discarded trash subsystem and a predefined reference state. This objective is preferable to direct state reconstruction fidelity, which seeks to maximize the similarity between the original input and the final reconstructed output, because it avoids the practical challenge of preparing multiple identical copies of the input state.

DQAS has been applied to quantum reinforcement learning, a subfield of quantum machine learning \cite{sun2023differentiable}. Within quantum reinforcement learning, variational quantum circuits serve as the counterpart to classical deep Q-networks, playing a critical role as the agent's decision-making model \cite{chen2020variational}. In operation, this quantum deep Q-network takes an observation state from the environment as input and encodes it into a quantum state. It subsequently predicts the expected future reward for each possible action by measuring the output state with an action-specific observable. During training, the network is optimized by encouraging these predictions to align with target values derived from future rewards.

Distributed QAS \cite{situ2024distributed} aims to design distributed quantum circuits for interconnected quantum processing units. This framework incorporates the assignment of the qubits in a quantum circuit to physical qubits across multiple quantum processing units into the search space. Simultaneously, it utilizes a virtual connectivity graph to manage the intricacies of nonlocal gate implementation, facilitating the integration of TeleGate and TeleData protocols \cite{caleffi2024distributed} directly into the design process. Through the coordination of multiple processing units, these methodologies enable the execution of complex computations across a greater number of qubits, providing a scalable pathway to tackle larger-scale problems.

Beyond the applications discussed above, QAS has also been applied to other tasks, including classification of quantum data \cite{lu2021markovian,su2025topology}, quantum generative adversarial networks \cite{duong2022quantum}, and variational quantum linear solvers \cite{wang2023automated}. These studies further demonstrate the versatility of QAS across different quantum computing scenarios. From a broader perspective on automated quantum design, closely related learning-based approaches have also been explored in other areas of quantum technology. For example, automated design approaches based on reinforcement learning have been explored for quantum communication, where learning agents can discover protocols such as quantum teleportation, entanglement purification, and quantum repeaters \cite{wallnofer2020machine}. Beyond protocol discovery, such approaches can also identify improved solutions for long-distance quantum communication, particularly in asymmetric settings with nonuniform channel noise and segment lengths. Reinforcement learning approaches have also been applied to quantum control to optimize control trajectories for fast and high-fidelity quantum gates \cite{niu2019universal}, and to quantum sensing to generate optimal quantum control signals for parameter estimation \cite{xiao2022parameter}. These related developments may also provide useful insights for extending QAS to a broader range of quantum technologies in the future.

\section{Emerging Trends}
\label{Sect:trends}

In this section, we provide a retrospective analysis of several key trends emerging in QAS that span different paradigms and indicate a transition toward more practical, adaptive, and specialized QAS methodologies.

A noticeable trend in recent QAS research is the transition from hardware-agnostic  search toward hardware-aware search. The circuit structure space is commonly constrained by the native gate set and qubit connectivity of the target device, while task performance is optimized within this hardware-compatible space. Hardware noise, which can degrade circuit fidelity and overall performance, is also increasingly incorporated into the QAS process. When evaluating candidate circuits, their performance is often assessed under noisy conditions, usually using quantum circuit simulators with noise models calibrated from real quantum devices. 
Under hardware noise, increasing circuit complexity does not necessarily improve task performance, since performance is jointly affected by circuit expressiveness and accumulated noise. Consequently, optimizing task performance under noisy conditions implicitly favors circuit structures that strike an appropriate balance between sufficient expressiveness and limited noise accumulation. 
Since noisy circuit simulation is substantially more expensive than noiseless simulation, alternative approaches decouple the evaluation of task performance and noise robustness. For example, Clifford noise resilience \cite{anagolum2024elivagar}, a training-free proxy, can estimate a circuit's robustness to hardware noise using only a small number of noisy circuit simulations. This proxy can be used either to filter out low-fidelity circuits or to be combined with a low-cost estimate of task performance in multi-objective optimization, thereby guiding the search toward circuits that are effective for the target task and robust to hardware noise.

Multi-objective optimization has become increasingly relevant in QAS because practical circuit design often involves several competing objectives rather than task performance alone. These objectives can be incorporated into the search through weighted combinations, constrained optimization, or Pareto-based approaches, to achieve an appropriate balance among competing requirements. 
In addition to the case discussed above, where noise robustness and task performance are evaluated separately using competing low-cost proxies, multi-objective QAS can also jointly optimize task performance and resource cost. For example, Ref. \cite{altares2021automatic} considers maximizing task accuracy while minimizing circuit size. Considering resource cost also helps improve the practical deployability of the searched circuits under limited quantum computing resources.

Recent QAS studies also show increasing methodological hybridization, moving beyond individual search methods toward combinations that leverage the complementary strengths of different methods. For example, Bayesian optimization has been combined with evolutionary strategies to optimize the acquisition function \cite{duong2022quantum}, while Monte Carlo tree search has been formulated with a combinatorial multi-armed bandit model to characterize the selection of unitary operators \cite{wang2023automated}. Meanwhile, various deep learning techniques have been introduced into QAS to improve representation learning, performance estimation, and search efficiency, including self-supervised learning \cite{he2023gsqas}, semi-supervised learning \cite{situ2026data}, meta-learning \cite{he2022quantum}, and mixture of experts \cite{he2025adaptive}. These developments provide greater flexibility in addressing the diverse challenges of quantum circuit design.

\section{Potential Directions}
\label{Sect:conclusion}

QAS has made rapid progress in recent years, driven by the need to design efficient and task-specific circuits under the constraints of near-term quantum computers. Despite these advances, many challenges remain, and there is significant room for further improvement. In this section, we outline several potential research directions that may help advance the field.

A significant limitation of current QAS methods is their restricted scale, as most existing studies are evaluated on systems with only a few to a dozen qubits. This constraint is primarily due to the fact that academic research teams often face difficulties performing QAS on physical quantum hardware because of limited access and high costs compared to major hardware providers. Consequently, many researchers rely on classical simulators where the exponential increase in required computational resources restricts the scale of experiments. However, current quantum hardware already features more than a hundred physical qubits \cite{kim2023evidence,google2025quantum} and this number is expected to grow rapidly. Furthermore, public access to quantum cloud platforms is expected to become more readily available in the near future. Achieving scalability is therefore a necessary direction for future research to ensure these QAS methods can be applied to systems with much higher qubit counts. One potential approach involves the use of modular or hierarchical search spaces that decompose large circuits into smaller and more manageable components. Another promising direction is low-dimensional embedding, which maps the original high-dimensional circuit search space into a compact continuous latent space where optimization can be performed more efficiently \cite{he2024gradient,sun2026quantum}. Additionally, some studies have improved scalability by warm-starting the QAS process with a matrix product state approximation of the target solution \cite{kundu2026tensorrlqas}, thereby providing a more favorable initialization for large-scale optimization. Exploring transfer learning could also enable the application of structural patterns discovered in small-scale systems to much larger systems.

A critical future direction for QAS is the enhancement of interpretability, which encompasses both circuit design and the analytical tools used to explain circuit performance. In terms of design, a strategic priority is to establish intrinsic interpretability by developing self-interpretable circuit structures. Rather than allowing the automated process to generate arbitrarily complex gate sequences, this approach restricts the search to circuit structures that adhere to specific physical symmetries or predefined theoretical principles. By combining high performance with a logical and transparent structure, circuits produced under these constraints can be readily analyzed by experts. Such inherent transparency improves the reliability of the resulting models, which is essential for their adoption in scientific research and real-world applications. Complementing this direction, there is also a growing need for post-hoc interpretability to better understand complex, high-performing quantum circuits. While searched structures may remain inherently opaque, post-hoc diagnostic tools can provide valuable explanations for their observed behavior. Future work should focus on methods that identify the most influential gates, subcircuits, or entanglement patterns, thereby offering deeper insights into how a circuit achieves its performance. By leveraging such external explanations, researchers can build greater trust in QAS-generated solutions, even when the underlying circuits remain highly sophisticated.

The performance of quantum circuits is closely linked to their underlying hardware. Most current methods in QAS assume a fixed hardware configuration with set qubit connectivity and static calibrated noise patterns. However, quantum hardware architectures are still rapidly evolving and have not yet converged on standardized designs. Consequently, treating the device configuration as a constant prevents the discovery of truly optimal system designs. A major research direction for the future is the joint optimization of quantum circuit structures and physical hardware configurations. This approach involves expanding the search space to include hardware parameters such as qubit placement and available coupling paths. By exploring a unified search space that includes both quantum circuit structures and hardware configurations, it is possible to discover synergistic designs where the circuit and the physical device are mutually optimized for maximum efficiency.

Current QAS approaches are largely derived from established discrete optimization methods and neural architecture search algorithms. While these approaches provide a functional starting point, they often overlook the fundamental differences between variational quantum circuits and classical neural networks. Quantum systems are governed by principles such as entanglement and interference which have no direct classical counterparts. Furthermore, constraints like the no-cloning theorem and the presence of hardware noise significantly influence the operational behavior and optimization methodologies of quantum circuits, distinguishing them from classical neural networks. Phenomena such as barren plateaus and restricted trainability also present unique scaling challenges that are distinct from the gradient issues encountered in classical deep learning. Consequently, future developments in this field must transition toward quantum-native search strategies that explicitly incorporate these physical characteristics into QAS.

Standardized benchmarking and reproducibility also deserve greater attention. At present, different studies often adopt different search spaces, target tasks and noise models, making direct and fair comparison difficult. In classical neural architecture search, widely used benchmarks usually specify both the search space and the target tasks, allowing different search algorithms to be evaluated under controlled and repeatable settings. Similar benchmarks could be developed for QAS by defining representative search spaces, target applications, noise model and evaluation metrics. The availability of open-source implementations would further improve the reproducibility and comparability of QAS research. These efforts would provide a more reliable basis for evaluating progress, comparing different QAS methods, and assessing their generalization across different search spaces, tasks, and hardware settings.

In conclusion, QAS provides a promising direction for automating quantum circuit design, though many challenges remain. Continued progress along the directions outlined above is expected to further enhance its effectiveness and broaden its applicability in future quantum computing systems.



\bibliographystyle{abbrv}
\bibliography{reference}

\end{document}